%% file: Galina_Patrick_2col.tex
\documentclass[conference]{IEEEtran}

\usepackage{cite}
\usepackage{ifthen}
\usepackage[T1]{fontenc}
\usepackage{enumerate}
\usepackage[american]{babel}
\usepackage{ctable}
\usepackage{graphicx}

\usepackage{xspace}
\usepackage{subfigure}
\usepackage{manfnt}
\usepackage{amstext,amsmath,amssymb}

\usepackage{pgf}
\usepackage{pgfplots}
\usepackage{tikz}
\usepackage{tikz-inet}
\usetikzlibrary{shapes,arrows,automata}
\usepackage{url}

\newtheorem{proposition}{Proposition}

\newcommand\define{:=}

\gdef\dd{\mathrm d}

\gdef\teq{\tilde t}
\gdef\Expmueta{\frac{1-e^{-kM}}{M^{1-v}N^{1-w}}}

\pgfcreateplotcyclelist{\mylist}{
	{red,mark=none,thick,densely dashed}, 
	{blue,mark=none,thick}, 
	{red,mark=none,thick,densely dotted}, 
	{mark=none,thick,dash pattern=on 2pt off 2pt on 6pt off 2pt}, 
}
\gdef\figwidth{4.85cm}
\gdef\figheight{4.6cm}

\newif\ifnotes\notestrue
\def\hpat#1{}

\newif\ifshort
\shortfalse

\pgfplotsset{compat=1.3}
\begin{document}
\newcounter{countenumeratep}

\title{Content Providers Volunteering to Pay Network Providers: Better than Neutrality?}

\author{
\IEEEauthorblockN{Patrick Maill\'e}
\IEEEauthorblockA{Institut Mines-Telecom\\
Rennes, France\\
Email: patrick.maille@telecom-bretagne.eu}
\and
\IEEEauthorblockN{Galina Schwartz}
\IEEEauthorblockA{UC Berkeley\\
USA\\
Email: schwartz@berkeley.edu}
}

\maketitle

\begin{abstract}
This paper studies the effects on user welfare of imposing network neutrality, using a game-theoretic model of provider interactions based on a two-sided market framework: we assume that the platform--the last-mile access providers (ISPs)--are monopolists, and consider content providers (CPs) entry decisions. All decisions affect the choices made by users, who are sensitive both to CP and ISP investments (in content creation and quality-of-service, respectively). 
In a non-neutral regime, CPs and ISPs can charge each other, while such charges are prohibited in the neutral regime. We assume those charges (if any) are chosen by CPs, a direction rarely considered in the literature, where they are assumed fixed by ISPs. 

Our analysis suggests that, unexpectedly, more CPs enter the market in a non-neutral regime where they pay ISPs, than without such payments. Additionally, in this case ISPs tend to invest more than in the neutral regime. 
From our results, the best regime in terms of user welfare is parameter dependent, calling for caution in designing neutrality regulations.
\end{abstract}

\IEEEpeerreviewmaketitle

\section{Introduction}

The network neutrality debate has been active for more than a decade, and is still vivid, as illustrated by the recent political decisions and interventions in Europe and the United States (e.g., B. Obama's intervention in Nov. 2014, FCC's February 2015 decision to regulate broadband services under Title 2, European Parliament's April 2014 report on neutrality).
The debate emerged from disputes between ISPs--carrying traffic but making fixed revenues from users--and CPs--making increasing revenue and attracting users with bandwidth-consuming services: ISPs' main claim was about charging CPs for bandwidth a so-called \emph{side payment} to cover their soaring costs in network upgrades and maintenance. Such practices have been labeled as \emph{non-neutral}, since opening the way to differentiated treatment of traffic based on economic considerations.
Much academic research has been carried out on the subject, investigating the consequences of regulating monetary exchanges between CPs and ISPs, on innovation (no entry barriers for CPs), user Quality of Experience, or user welfare (see~\cite{wu2003,lenard2006net,WuAER2009,maille2014telecommunication} and references therein). When side-payments are considered, they are assumed fixed by ISPs. On the contrary, in this paper we investigate the consequences of letting the side-payment decision to CPs, even allowing CPs to choose negative prices (i.e., to get paid by ISPs). Our analysis shows that when actors are rational, this can lead to satisfying situations: CPs indeed need ISPs to attract users, and hence will care about ISP survivability when selecting the side-payment values.

We develop a game-theoretic model in which the network regime (neutral or non-neutral) affects strategic provider choices, such as: investments of last-mile access providers (ISPs) and content providers (CPs); ISP pricing of their end-users; and CPs entry decisions and choices of side payments to the ISP(s). 
We model CP-ISP interactions using a two-sided market framework~\cite{RoTi2006, Armstrong, RysmanAER2009}. 

A number of researchers
applied the ideas of two-sided markets to study network neutrality. Hermalin
and Katz \cite{HK2006} model neutrality as a restriction on the
product space, and consider whether ISPs should be allowed to offer differentiated services. Hogendorn \cite{Hogendorn} studies two-sided markets
where intermediaries sit between \textquotedblleft conduits\textquotedblright%
\ and CPs. In his context, neutrality means that content has open access
 to the intermediaries. Njoroge \emph{et al.}~\cite{Ozdaglar2013} consider a two-sided market
model with heterogeneous CPs and end-users, and the ISPs play the role of a
platform. Weiser \cite{Weiser}, and Lee and Wu \cite{WuAER2009} discuss policy issues related to two-sided markets.
 Schwartz \emph{et~al.}~\cite{schwartz2012} study the effects of network neutrality on ISP and CP entry.
 Since network regime profitability affects provider profits, the market structure is affected as well. 
Remark that, for realistic parameters,  in most two-sided market models, aggregate user welfare is higher in a non-neutral regime, see for example \cite{MSW_neutrality_09, schwartz2012, EconomidesTag2012, Ozdaglar2013}.
  While our model closely relates to these papers,  here CPs choose their payment to ISPs. As a result, the analysis of the game is considerably different: instead of 3 stages in \cite{schwartz2012}, our game has five stages since three strategic decisions previously made by ISPs (hence, simultaneously) are now split between ISPs and CPs (hence sequentially).

We model a non-neutral network as allowing side payments between CPs and ISPs, and a neutral one as forbidding such side payments. Our end-user demand for content variety has a flavor of the classical monopolistic competition model \cite{DS1977}.
CPs derive their revenue from advertisers, and ISPs from charging end-users for traffic; in addition, in the non-neutral regime, side payments affect provider revenues.
We consider a usage-based (actually, pay-per-click) pricing scheme for users, as well as a price per click for CPs who (in the non-neutral regime) pay the ISP for delivering content to users. While~\cite{MSW_neutrality_09} assumes both of those prices are chosen by the ISP to maximize its profit, here we propose that the CPs ``volunteer'' to pay to ISPs by deciding the value of the side-payment. This gives rise to a multi-level game that we analyze by backward induction to find an equilibrium. Numerical results indicate that it can be in the best interest of CPs to voluntarily choose to pay the ISPs, and that this can result in a higher aggregate end user welfare than the no-side-payment situation.

\emph{Ceteris paribus}, a neutral regime tends to be better for aggregate user welfare when advertising rates are low, price sensitivity of end-user demand is high, and added content variety (modeled as a higher number of CPs) has little effect on user demand, and CP entry costs are low.
In reverse, a non-neutral regime performs better when user demand has low sensitivity to price and high sensitivity to content variety, and CP entry costs are high. 

The remainder of the paper is organized as follows: Section~\ref{sec:model} introduces the setting and the mathematical model we consider, and we present and compare the analytical resolution for both neutrality scenarios in Section~\ref{sec:analysis}. We then introduce and discuss the CP entry game in Section~\ref{sec:CPentry}, and provide a numerical analysis in Section~\ref{sec:numerical}. Conclusions and directions for future work are given in Section~\ref{sec:conclusion}.

\section{Model}\label{sec:model}

Let us assume that $N$ (the number of ISPs) and $M$ (the number of CPs) are fixed. (Later on, we will consider an entry game for CPs.)
 Throughout the paper we refer to a particular ISP through the letter $n$, and a particular CP through the letter $m$.

\subsection{Payment structure}

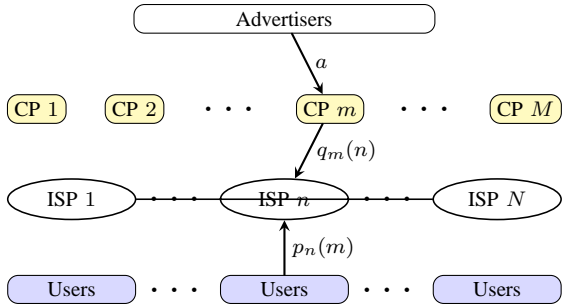
\begin{figure}[htbp]
\centering
{\footnotesize
\begin{tikzpicture}[line cap=round,line join=round,>=triangle 45,x=1.0cm,y=1.0cm,yscale=.3,xscale=.6,node distance=1.3cm]
\newdimen\mydimy
\node (monnoeud) at (1,1) {};
\pgfextracty{\mydimy}{monnoeud}

\tikzstyle{ISP}=[ellipse,draw,semithick,minimum height=\mydimy,minimum width=1.7cm]
\tikzstyle{CP}=[rectangle,draw, minimum size=\mydimy, rounded corners=4pt,fill=yellow!30]
\tikzstyle{User}=[rectangle,draw, minimum width=1.7cm, rounded corners=4pt,fill=blue!15]
\tikzstyle{suite}=[->,>=stealth,thick,rounded corners=4pt]
\node[CP] (CP1) at (0,0) {CP $1$};
\node[CP,right of=CP1](CP2){CP $2$};
\node[right of=CP2](CPdots1){{\LARGE $\dots$}};
\node[CP,right of=CPdots1](CPm){CP $m$};
\node[right of=CPm](CPdots2){{\LARGE $\dots$}};
\node[CP,right of=CPdots2](CPM){CP $M$};

\node[ISP] (ISP1) at (0,-4-|CP1.west)[anchor=west] {ISP $1$};
\node[ISP] (ISPN) at (ISP1-|CPM.east)[anchor=east] {ISP $N$};
\draw[opacity=0](ISP1)--node(ISPn)[ISP,opacity=1]{ISP $n$}(ISPN);
\draw[opacity=0](ISP1)--node[opacity=1]{{\LARGE $\dots$}}(ISPn);
\draw[opacity=0](ISPN)--node[opacity=1]{{\LARGE $\dots$}}(ISPn);

\draw (ISP1)++(0,-4) node(User1)[User]  {Users};
\node[User] (Usern) at (User1-|ISPn) {Users};
\node[User] (UserN) at (User1-|ISPN) {Users};
\draw[opacity=0](User1)--node[opacity=1]{{\LARGE $\dots$}}(Usern);
\draw[opacity=0](UserN)--node[opacity=1]{{\LARGE $\dots$}}(Usern);

\node[rectangle,draw,minimum width=4cm,rounded corners=4pt] (Adv) at (0,4-|ISPn){Advertisers};

\draw[suite] (Adv) -- node[anchor=west]{$a$} (CPm);
\draw[suite] (CPm) --node[anchor=west]{$q_m(n)$} (ISPn);
\draw[suite]  (Usern) -- node[anchor=west]{$p_n(m)$} (ISPn);
\end{tikzpicture}
}
\caption{Payments for each click of an ISP $n$ user to get CP $m$ content ($q_m(n)=0$ in the neutral regime). The strategic players are ISPs--setting prices $(p_n(m))$ and investment levels $(t_n)$--and CPs--setting prices $(q_m(n))$ and investment levels $(c_m)$.}
\label{fig:payments}
\end{figure}

Payments are illustrated in Figure~\ref{fig:payments}: each ISP acts as a monopolist over some population of users (subscribers), which are charged per usage. On the other side of the market, CPs also pay the ISP (in the non-neutral regime) to access users.
More precisely, if an ISP $n$ user clicks to obtain some content from a CP $m$, that user pays $p_n(m)$ to her ISP and the CP pays $q_m(n)$ to that ISP. CPs make money through advertising: we assume each click earns the CP a fixed amount $a$.

\subsection{ISP profits}

The number of clicks $B_{n,m}$ of ISP $n$ users for CP $m$ content is assumed to be as in~\cite{MSW_neutrality_09}:
\begin{equation}\label{eq:expression_Bnm}
B_{n,m}=\gamma c_m^v t_n^we^{-p_n(m)/\theta};
\end{equation}
with $\gamma=\Expmueta$ reflecting the user preferences for content diversity\footnote{The denominator in $\gamma$ acts as a normalization factor.} in the game with given $M$ and $N$, and the other variables explained in Table~\ref{tab:variables}%
.
\begin{table}[htbp]
\begin{center}
\begin{tabular}{c|p{.39\textwidth}}
$k$& user preference for content variety\\
$c_m$& CP $m$ investment (content attracts users)\\
$v$& demand sensitivity to CP investments\\
$t_n$& ISP $n$ investment (attracts user demand)\\
$w$& demand sensitivity to ISP investments\\
$p_n(m)$ & price (per click for the $m$-th CP content) payed to the $n$-th ISP by its users\\
$q_m(n)$ & price (per click of the $m$-th CP content for $n$-th ISP users) payed by the $m$-th CP to the $n$-th ISP \\
$\theta$& user insensitivity to price\\
$\alpha>1$&outside option for ISPs\\
$\beta>1$&outside option for CPs\\
$c_e$&CP entry cost
\end{tabular}
\caption{Parameters and variables of the model}
\label{tab:variables}
\end{center}
\end{table}

Although quite complex, the expression~\eqref{eq:expression_Bnm} intends to cover the main factors affecting demand: the number of clicks from ISP $n$ users on CP $m$ content depends on content diversity ($M$), quality of service (through ISP $n$ investments), content quality (through CP $m$ investments), and the click-price $p(n)$. The exponential form for the price-related part can be interpreted as stemming from an exponential distribution of the willingness-to-pay among users; for the investment-related parts we use a Cobb-Douglas utility formulation, which is classically used to model the preferences among complementary resources (here, in content and network quality which are complementary to attract demand)~\cite{varian1992microeconomic}. Note that we assume $v+w<1$, i.e., decreasing returns to scale.

The total number of clicks for ISP $n$ is 
\[
B_n=\gamma(c_1^v+...+c_M^v)t_n^we^{-p_n(m)/\theta},
\]
we can therefore express the profit of ISP $n$ as
\begin{eqnarray}
\Pi_{T_n}&\!\!\!=\!\!\!&\sum_{m=1}^M(p_n(m)+q_m(n))B_{n,m}-\alpha t_n\nonumber\\
&\!\!\!=\!\!\!&\sum_{m=1}^M(p_n(m)+q_m(n))\gamma c_m^vt_n^we^{-p_n(m)/\theta}-\alpha t_n,\label{eq:ISPprofit}
\end{eqnarray}
where $q_n(m)$ is a price paid by CP $m$ for each click by a customer of ISP $n$ (the so-called side-payment, that may be imposed by a regulator) and $\alpha>1$ is the outside option (gain from alternative uses of funds).

\subsection{CP profits}

CPs make profit through advertising (hence the incentive to attract clicks) and undergo the side payment to ISPs.
The profit of each CP $m, 1\leq m\leq M$ therefore equals
\begin{eqnarray}
\Pi_{C_m}&=&\sum_{n=1}^N B_{n,m}(a-q_m(n))- \beta (c_m +c_e)\nonumber\\
&=&c_m^v \gamma \sum_{n=1}^N(a-q_m(n))t_n^we^{-p_n/\theta}  - \beta (c_m +c_e),\label{eq:CPprofit}
\end{eqnarray}
with $a$ the average revenue per click from advertisements, 
$\beta>1$ the outside option and $c_e$ the CP entry cost.

\subsection{Order of decisions}

In this paper we consider two regimes, that lead to different values of investments and prices.

\subsubsection{Neutral regime}

Here, the side payments $(q_m(n))_{1\leq m\leq M,1\leq n\leq N}$ are imposed to be $0$. Decisions are made in the following order:
\renewcommand\theenumi{\alph{enumi}}
\begin{enumerate}
\item at the largest time scale, the ISPs decide how much to invest (variable $t_n$ for each $n$) and how much to price each click (variables $(p_n(m))_{1\leq m \leq M}$ for each ISP $n$), through a \emph{non-cooperative game among ISPs}.
\item at a smaller time scale, the CPs decide how much to invest (variable $c_m$ for each $m$). 
\end{enumerate}
That regime has been studied in~\cite{MSW_neutrality_09}.

\subsubsection{Non-neutral regime}
In that regime, the side payments also have to be decided. In this paper we assume they are chosen by CPs, and that this decision takes place after ISP investment decisions but before ISP pricing decisions. Indeed, we expect that ISP investments (infrastructure costs) are for large time scales, due to the enormous economic stakes, and the fact that such investments correspond to time-consuming actions (adding underground and/or underwater cables, possibly renegotiating agreements with backbone providers...). On the other hand, the prices applied to users can change quite rapidly. We consider that side-payments are set through contracts (bargaining) between ISPs and CPs, which take place at an intermediate time scale between those two. We keep the CP investments for the smaller time scale, since content creation can be very responsive to the conditions (i.e., to the prices and side-payment levels as well as ISP investments). Summarizing, the game is played as follows (from the largest time scale to the smallest):
\begin{enumerate}
\item ISPs decide their investments $(t_m)_{1\leq m\leq M}$
\item Each CP $m$ chooses what (possibly negative) price $q_m(n)$ to pay to each ISP $n$
\item ISPs set their prices $(p_n(m))_{1\leq n\leq N,1\leq m\leq M}$
\item CPs set their investments $(c_m)_{1\leq m\leq M}$
\end{enumerate}
Note that the order of decisions is the same for both the neutral and non-neutral regimes: in the neutral regime the side-payment choice is absent hence ISPs can simultaneously decide their investments and prices.
Of course this order of decisions can be discussed, and variants are worth considering. 

\section{Analysis of the games}\label{sec:analysis}

We analyze the (Stackelberg) interactions among actors by backward induction~\cite{fudenberg1991game}, meaning that decisions at each stage are made anticipating the consequences on the smaller-time-scale stages.

\gdef\tneutral{t^{\text{neut}}}
\gdef\pneutral{p^{\text{neut}}}
\gdef\cneutral{c^{\text{neut}}}
The outcome for the neutral case can be found in~\cite{MSW_neutrality_09}. We recall that result here.
\begin{proposition}\label{prop:neutral_case}
The neutral game has a unique equilibrium, where
\begin{itemize}
\item all ISPs charge their users the same price per click
\[
\pneutral = \theta\pi \qquad \text{ where } \pi\define \frac{1}{1+\frac{v}{N(1-v)}};
\]
\item investments from ISPs all equal
\begin{equation}
\tneutral =\left[\left(x\right)^{1-v}\left(\frac{av}{\beta}\right)^{v}e^{-\pneutral/\theta}\right]^{\frac{1}{1-v-w}}
\end{equation}
where $x\define(1-e^{-kM})^{\frac{1}{1-v}}\cdot\left(\frac{\theta w}{\alpha}\right)N^{\frac{v+w-1}{1-v}}$;
\item all CPs invest at the same level
\[
\!\!\!\cneutral = \!\!\left[\left(x\right)^{w}\!\left(\!\frac{av}{\beta}\!\right)^{1-w}\!\!\!\!e^{-\pneutral/\theta}\right]^{\frac{1}{1-v-w}}\!\!\!\!\cdot \left(\!\frac{1-e^{-kM}}{M^{1-v}} N^w\!\!\right)^{\!\!\frac{1}{1-v}}\!\!\!. 
\]
\end{itemize}
\end{proposition}

\begin{IEEEproof}
See~\cite{MSW_neutrality_09}.
\end{IEEEproof}

For the non-neutral case, we obtain the following result.
\begin{proposition}\label{prop:the_game}
The non-neutral game has a unique symmetric equilibrium, where 
\begin{itemize}
\item investments from ISPs all equal
\begin{equation}\label{eq:t_eq_final}
\teq=\left(\left[(1-R)\frac{M}{\tilde K}\frac{w}{1-v} \right]^{1-v}\theta e^{q/\theta-\pi} \pi (1-\pi)^v\right)^{\frac{1}{1-v-w}}
\end{equation}
\vspace{-2ex}
\[
\hspace{-9ex}\text{where~~}\left\{
\begin{array}{lcl}
\pi&=&\frac{1}{1+\frac{v}{N(1-v)}}\\
R&=&\frac{N-1}{N}\frac{\frac{1}{\pi}+\frac{v}{1-\pi}-1}{\frac{1}{\pi}+\frac{1}{1-\pi}-1}\\
\tilde K&=&\frac{\alpha}{\gamma}\left(\frac{v}{1-v}  \frac{v\gamma }{\beta}\right)^{-\frac{v}{1-v}}.
\end{array}
\right.
\]
\item payments from CPs to ISPs are all equal, to 
\begin{equation}
q_m(n)\define q=a-\theta;
\end{equation}
\item user prices set by ISPs equal
\[
p_n(m)=\tilde p=\theta\pi-q
=\theta(1+\pi)-a.
\]
\item CP invest to a level 
\begin{eqnarray}
c=\left(  \frac{v\gamma}{\beta}N\theta \teq^{w}e^{a/\theta-(1+\pi)}\right)^{\frac{1}{1-v}},\label{eq:c_final}
\end{eqnarray}
where we do not replace $\teq$ by its expression for space reasons.
\end{itemize}
\end{proposition}

\begin{IEEEproof}
\ifshort
\else
{The proof is provided in the Appendix. It follows the well-known backward induction method: using first-order optimality conditions, we first express the CP investment levels as a function of the other decisions (prices and investments); then we include that decision in the ISP objective to compute their optimal prices. Expressing the optimal side-payments and then ISP investment involve quite some algebra, but the methodology is the same.
}
\fi
\end{IEEEproof}

\gdef\UW{\text{UW}}

The analytical expressions for investments obtained for both cases are quite complex, and difficult to interpret. However, the \emph{prices} imposed to end-users, and the side-payments, are very simple and can be interpreted as follows:
\begin{enumerate}
\item in the non-neutral regime, side payments are simply the difference between the advertisement revenue per click and the average user willingness-to-pay for a click (with our interpretation of willingness-to-pay values to be exponentially distributed with mean $\theta$). If users are willing to pay more than what advertising yields, ISPs are making ``too much'' and the side-payment balances things out. Reciprocally, if advertising revenues are high with respect to what users are willing to pay, CPs \emph{choose} to compensate ISPs for each click, so that ISPs later decide on low click-prices and hence high demand, which in the end benefits to CPs in the form of advertising revenues. The latter case is particularly interesting, since it illustrates how rational CPs may decide to support ISPs via side payments. 
\item Now let us have a look at the price paid by users: with respect to the neutral regime, we observe that ISPs end up entirely reporting the side payments from CPs onto users. When those side-payments are positive (CPs paying ISPs), users are therefore charged less than in the neutral regime. As from the above point, this is the case when advertising revenues exceed the average willingness-to-pay: to globally make more revenue (from advertising), ISPs ``subsidize'' users by reducing their click-prices. We also have the opposite effect when $a<\theta$: users being ``rich'' and yielding low ad revenues, ISPs decide to charge them more.  
\end{enumerate}

But if we want to compare both situations at a global scale, prices are not the only results to look at since \emph{investments} affect the quality of experience, the demand, and (in the CP entry game) the number of CPs that enter the market.
To have a reasonable comparison metric, we define the \emph{user surplus} $\UW$ at the equilibrium, as the cumulated value of the system for users, i.e.,
\begin{eqnarray*}
\UW &=& \sum_{n=1}^N\sum_{m=1}^M\int_{p=\tilde p}^{+\infty} B_{n,m}(p) \dd p\\
&=& NM \gamma\theta {\tilde c}^v \teq^w e^{-\tilde p/\theta}
\end{eqnarray*}

Also for comparison purposes, let us express the revenue of each ISP $n$ (still at equilibrium) for the non-neutral regime:
\begin{eqnarray*}
\Pi_{T_n}&\!\!\!\!\!=\!\!\!\!\!&M\theta\gamma\left(\theta \frac{v}{1-v}  \frac{v\gamma }{\beta}\right)^{\frac{v}{1-v}} \teq^{\frac{w}{1-v}} e^{\frac{q/\theta-\pi}{1-v}}\pi^{\frac{1}{1-v}}(1-\pi)^{-\frac{v}{1-v}}
\\&& -\alpha \teq,
\end{eqnarray*}
with $\teq$ given in~\eqref{eq:t_eq_final}. Similarly one can express the ISP revenue in the neutral regime.

Focusing on the CP point of view, the revenue of each CP $m$ in the non-neutral regime is
\begin{equation}\label{eq:CP_revenue_eq}
\Pi_{C_m}=\left(N(a-q)\frac{\gamma}{(\beta/v)^v} \right)^{\frac{1}{1-v}}\teq^{\frac{w}{1-v}}e^{\frac{q/\theta-\pi}{1-v}}(1-v)-\beta c_e,
\end{equation}
and the one for the neutral regime can similarly be drawn from Proposition~\ref{prop:neutral_case}.

Again, those expressions are too complex to allow direct interpretations. For that reason, in Section~\ref{sec:numerical} we will use numerical results on a given scenario to compare both neutrality regimes.

\section{Entry game for Content Providers}\label{sec:CPentry}

Until now, we had assumed that the numbers of CPs and ISPs were fixed. We now relax that assumption for CPs, by considering that new CPs enter the market as long as it is profitable for them, i.e., as long as the revenue~\eqref{eq:CP_revenue_eq} with one additional CP remains positive. On the other hand, the number of ISPs is still assumed fixed.
We then have the following result.

\gdef\Mopt{{\cal M}}
\begin{proposition}\label{prop:uniqueM_fixedN}
Assume that the CP entry cost $c_e$ is strictly positive, and denote by $\Pi_{C_m}(N,M)$ the revenue of each CP at the symmetric equilibrium of the investments/prices game $G(N,M)$ with $N$ ISPs and $M$ CPs. 
Then for any fixed number $N$ of ISPs, the number $M$ of CPs that enter the game is uniquely defined.

More precisely, let us define $\Mopt$ as the unique strictly positive solution $x$ of $x=\frac{1-v-w}{k}\left(e^{kx}-1\right)$. Then,
\begin{itemize}
\item either
$
\Pi_{C_m}(N,M)< 0 \mbox{ for } M=\lfloor\Mopt\rfloor \mbox{ and }M=\lceil\Mopt\rceil, 
$
in which case $\Pi_{C_m}(N,M)< 0$ for any $M\geq 1$, and no CP enters the game;
\item or 
$
\{M\geq 1: \Pi_{C_m}(N,M)\geq 0\mbox{ and }\Pi_{C_m}(N,M+1)< 0\}
$
is a singleton, giving the number of CPs entering the game. In particular, that number is larger than $\lfloor\Mopt\rfloor$.
\end{itemize}
\end{proposition}

\begin{IEEEproof}
We focus on the dependency in $M$ of $\Pi_{C_m}(N,M)$ at the equilibrium of $G(N,M)$. \ifshort\else{From~\eqref{eq:cm_opt} and~\eqref{eq:PiCm_Am},}\fi{} the variations (sign of the derivative with respect to $M$, seen as a continuous variable) of $\Pi_{C_m}(N,M)$ are the same as those of the equilibrium CP investment $c$, itself being given in~\eqref{eq:c_final}, were the only term involving $M$ is $\gamma \teq^w$.

Isolating in~\eqref{eq:t_eq_final} the terms that depend on $M$ (namely, $\gamma$ and $M$), we conclude that the variations of $\gamma \teq^w$ (and hence, of $\Pi_{C_m}(N,M)$) follow those of $\gamma \left(M^{\frac{1-v}{1-v-w}}\gamma^{\frac{1}{1-v-w}}\right)^w=\left(\gamma M^w\right)^{\frac{1-v}{1-v-w}}$, that are simply those of 
$g(M)\define M^w\gamma=\frac{1-e^{-kM}}{M^{1-v-w}}$. The function $g$ is strictly positive on ${\mathbb R}^+$, tends to $0$ at zero, and its differential has the same sign as
\[x-\frac{1-v-w}{k}\left(e^{kx}-1\right),\]
which is concave, null at $0$, and tending to $-\infty$ as $x$ tends to infinity. Additionally it is strictly positive for $x>0$ sufficiently small (since having a strictly positive derivative at $0$) and therefore has a unique strictly positive root $\Mopt$: $g(x)$ is then strictly increasing for $x<\Mopt$ and strictly decreasing for $x>\Mopt$.

As a result, if CP revenues are strictly negative for the integers closest to $\Mopt$ then they are negative for any $M$, and otherwise at least $\Mopt$ CPs will enter the market, their number increasing until the CP revenue from an additional entry gets strictly negative.
\end{IEEEproof}

Proposition~\ref{prop:uniqueM_fixedN} provides us with a method to compute the unique number of CPs entering the game, which we will use in the next sections on a specific numerical example.

\section{Numerical results with an entry game for Content Providers}\label{sec:numerical}

We propose here an illustration of our results with the default parameter values displayed in Table~\ref{tab:parameter_values}, in order to compare the neutral and non-neutral settings with respect to the terms whose expressions were too complex to interpret, and which involved in particular the equilibrium investment values. We focus on the consequences of the CP-ISP interactions, specifically user welfare and paid prices, as well as the investment level quantified by the number $M$ of CPs entering the market.

Choosing ``real'' parameter values for our model is a difficult task, which would involve collecting and analyzing market data, and is beyond the scope of this paper. We take values that seem realistic to us, but are aware more work is needed to fine-tune those parameters in order to closely represent a specific market.%
\gdef\XN{2}
\begin{table}[htbp]
\begin{center}
{\footnotesize
\input{parameters.tex}
}
\caption{Default parameter values}
\label{tab:parameter_values}
\end{center}
\end{table}
With respect to our observations following Proposition~\ref{prop:the_game}, we are here in a case where advertisement revenues ($a$) exceed the average user willingness-to-pay ($\theta$): in the non-neutral regime CPs with therefore choose to pay a positive side-payment to ISPs, in order to stimulate demand (and thus advertisement gains).

\subsection{Influence of ad revenues and user average willingness-to-pay}

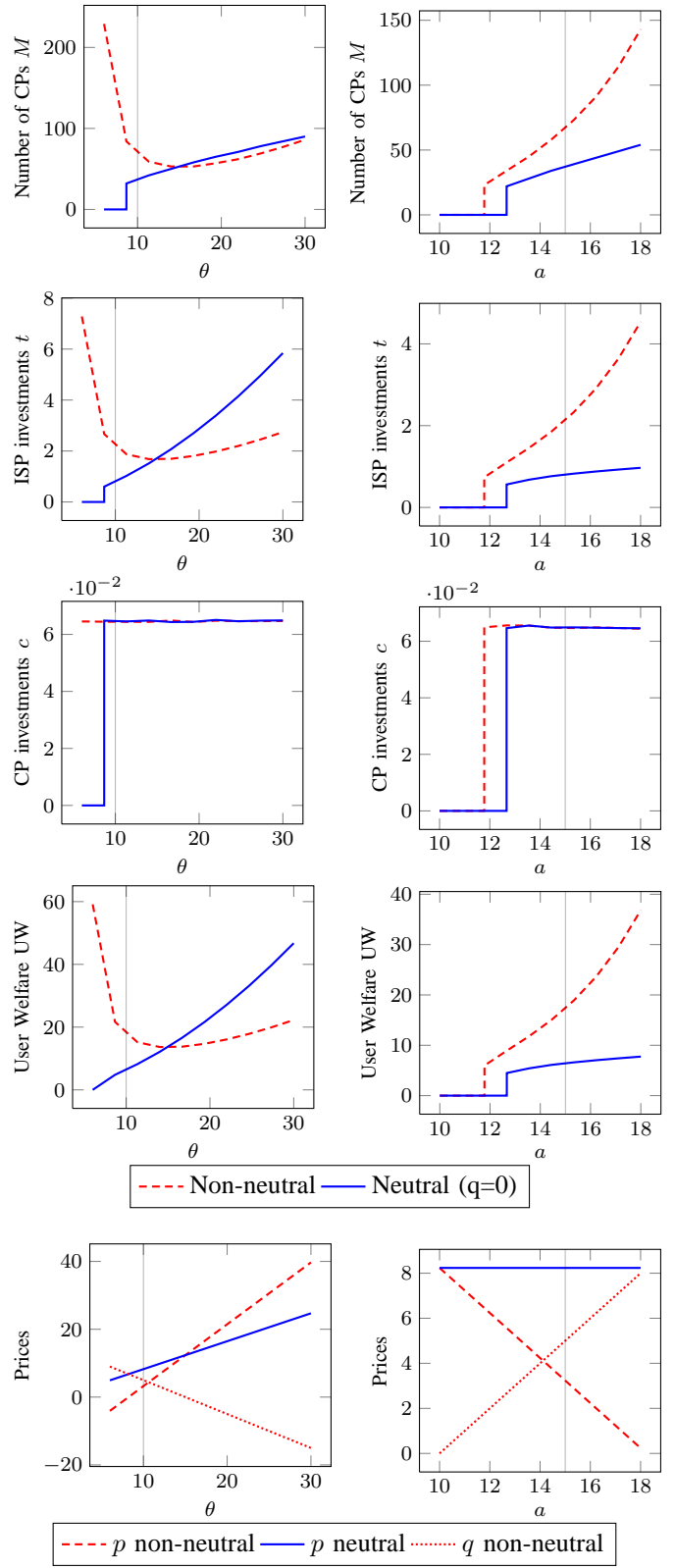
\begin{figure}[htbp]
\begin{center}
\input{M_fixedN_theta}
\hfill
\input{M_fixedN_a}
\input{t_fixedN_theta}
\hfill
\input{t_fixedN_a}
\input{c_fixedN_theta}
\hfill
\input{c_fixedN_a}
\input{UW_fixedN_theta}
\hfill
\input{UW_fixedN_a}
\\
\ref{named}
\vspace{3ex}\\
\input{p_fixedN_theta}
\hfill
\input{p_fixedN_a}
\\
\ref{priceleg}
\caption{Impact of user price insensitivity $\theta$ and advertising revenues $a$ on the number of CPs ($M$), ISP investments ($t$), CP individual investments ($c$), user welfare, and prices (including prices per click and side payments), with $N=\XN$ ISPs. A vertical line indicates the reference case (values in Table~\ref{tab:parameter_values}).}
\label{fig:M_UW_prices_vs_a_theta}
\end{center}
\end{figure}

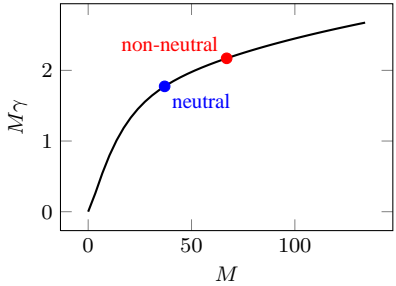
\begin{figure}[htbp]
\begin{center}
\input{Gamma_fixedN}
\caption{$M\times\gamma$ (quantifying how users value content diversity) versus $M$ for $N=\XN$}
\label{fig:MGamma_vs_M}
\end{center}
\end{figure}

Figure~\ref{fig:M_UW_prices_vs_a_theta} 
shows the evolution of the number $M$ of CPs entering the market, the CP and ISP investment levels and corresponding user welfare at equilibrium, and the different prices, when the user insensitivity to prices $\theta$ (which we can also interpret as the average user willingness-to-pay if CP and ISP investments were set to $1$) and the revenue-per-click from advertisers $a$ vary. 
Remark that in the neutral regime, when advertising does not yield enough revenue ($a$ too small), no CP enters the market and user welfare is null; on the contrary, in the non-neutral regime, negative side payments are sufficient to attract CPs in those cases. For our default values, the non-neutral regime attracts more CPs than the neutral one and leads to smaller prices per click for users, both effects contributing to a higher user welfare. In particular, we are in a zone where users value extra content diversity significantly, as illustrated by the slope of $M\gamma$ in the region of interest displayed in Figure~\ref{fig:MGamma_vs_M}. Remark also that CP investments seem to be approximately constant, as soon as entering the market is beneficial. 
\def\figwidth{6cm}
However, those effects are inverted when users become very price-insensitive (values of $\theta$ above $a$): then the neutral regime leads to more CPs, cheaper clicks for users, and more user welfare. Note that in this region the CPs actually get paid by ISPs since $q<0$. We nevertheless expect to rather be in the region when $a>\theta$, given the trend in the last decade toward advertisement-financed (rather than price-based) services.

Finally, Figure~\ref{fig:UW_vs_v_ce} highlights the influence of the parameter $v$ (demand sensitivity to content creation) and of the CP entry cost on user welfare. 
{\def\figwidth{4.5cm}
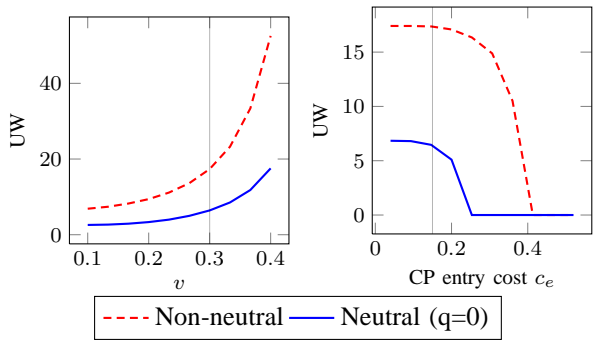
\begin{figure}[htbp]
\begin{center}
{\footnotesize
\input{UW_fixedN_v}
\input{UW_fixedN_c_e}
\\
}
\ref{named}
\caption{Impact on demand sensitivity to content $v$ and CP entry cost $c_e$ on user welfare.}
\label{fig:UW_vs_v_ce}
\end{center}
\end{figure}
As expected, large entry costs imply fewer CPs and thus reduce user welfare, but the non-neutral regime always performs better than the neutral one (still, with our values where $a$ exceeds $\theta$). A larger $v$ increases the difference (even the ratio) between the two regimes, in favor of the non-neutral one.

\section{Concluding Comments}\label{sec:conclusion}

From our results, the superiority of one regime over the other depends on parameters: advertising rates, price sensitivity of end-user demand,
CPs' entry costs, the magnitude of effect of the added content variety on user demand, and the relative importance of CPs and ISPs investments. We therefore suggest that our results call for caution in designing neutrality regulations.

However, the relative values of the expected revenues from advertisement and the average willingness-to-pay of users seem to play a particular role: when advertisement revenues are small, our model suggests that a neutral regime should be preferred, while with high advertisement revenues (with respect to user willingness-to-pay) a non-neutral regime leads to lower prices for users, higher user welfare and higher ISP investments. This situation is due to CPs volunteering to fund ISPs through side-payments, so as to stimulate their investments in order to earn CPs more advertising revenue through higher user demand. As we think we are currently in a situation where users tend to be willing to pay less and less, and many services are financed through advertising, such non-neutral settings are worth considering.

Among directions for future work, we think that considering competing ISPs (instead of here, several ISPs each being a monopolist over some population) deserves a thorough analysis. Also, in such a context it would be interesting to consider an entry game for ISPs as well. Finally, to carry out numerical analyses on some specific markets, some econometric work is needed to fine-tune the setting of the model parameters.

\bibliographystyle{plain}
\bibliography{NetNeut2015,biblio}

\ifshort
\else
{
\newpage
\appendix

\section{Proof of Proposition~1}

\subsection{CP optimal investment choice}

Assume that $N$ and $M$ are fixed, as well as prices and ISP investments. The revenue of CP $m,1\leq m\leq M$ is
\[
\Pi_{C_m}=\gamma c_{m}^{v}\sum_{n=1}^{N}(a-q_{n}(m))t_{n}^{w}%
e^{-p_{n}(m)/\theta}-\beta (c_{m}+c_e).
\]
The first-order condition for $c_{m}$ to be optimal to CP $m$ yields
\begin{equation}
c_{m}=\left(  \frac{v\gamma}{\beta}\sum_{n}(a-q_{m}(n))t_{n}^{w}%
e^{-p_{n}(m)/\theta}\right)  ^{\frac{1}{1-v}}. \label{e.a1c}%
\end{equation}
For notational convenience, let us define
\begin{equation}\label{eq:Am_def}
A_{m}:=\sum_{n}(a-q_{m}(n))t_{n}^{w}e^{-p_{n}(m)/\theta}
\end{equation}
so that\vspace{-5.5ex}
\begin{equation}\label{eq:cm_opt}
c_{m}=\left(  A_m\frac{v\gamma}{\beta}\right)  ^{\frac{1}{1-v}}.
\end{equation}%
The resulting profit $\Pi_{C_m}$ optimized with respect to $c_{m}$ is:
\begin{equation}\label{eq:PiCm_Am}
\Pi_{C_m}=\left(A_m\frac{\gamma}{(\beta/v)^v} \right)^{\frac{1}{1-v}}(1-v)-\beta c_e.
\end{equation}

\subsection{ISP choices of user prices}

Now consider ISPs fixing user prices $(p_n(m))_{1\leq m\leq M,1\leq n\leq N}$ through a noncooperative game, anticipating optimal CP investments.

Substituting (\ref{eq:cm_opt}) into \eqref{eq:ISPprofit}, ISP $n$ prices should maximize
\[
\sum_{m=1}^{M}\left\{  \gamma\left[  p_n(m)+q_m(n)\right]
t_{n}^{w}e^{-p_{n}(m)/\theta}\times\left[  \frac{v\gamma}{\beta}\right]
^{\frac{v}{1-v}}A_{m}^{\frac{v}{1-v}}\right\},
\]
where $A_{m}$ is given by
(\ref{eq:Am_def}). Considering $\mathbf{q}_{m}$ and $\mathbf{t}$ as fixed (already
chosen), and noting that $A_k$ is independent of $p_n(m)$ when $k\neq m$, each user price $p_n(m)$ should maximize
\[
(p_n(m)+q_m(n))e^{-p_{n}(m)/\theta}A_{m}^{\frac{v}{1-v}}.
\]
Since $\frac{\dd A_m}{\dd p_n(m)}=-\frac{a-q_m(n)}{\theta}t_n^we^{-p_n(m)/\theta}$, this leads to the first-order condition for each $n,1\leq n\leq N$ and $m,1\leq m\leq M$:
\begin{equation}
\frac{\theta}{p_n(m)+q_m(n)}-1=\frac{v}{1-v}\frac{(a-q_{m}(n))t_{n}^{w}e^{-p_{n}(m)/\theta}}{A_{m}}. 
\label{e.FOC_p_GS}%
\end{equation}
Summing over $n$, we obtain
\begin{equation}
\sum_{n=1}^{N}\frac{1}{p_n(m)+q_m(n)}=\frac{1}{\theta
}\left(  \frac{v}{1-v}+N\right). \label{e.q+p_nosymmetry}%
\end{equation}

\subsection{CP optimal choice of $q_{m}(n)$ \label{sec:CPoptQ}}

From \eqref{eq:PiCm_Am}, we
observe that maximizing $\Pi_{C_m}$ with respect to $q_{m}(n)$ is identical to maximizing
$A_{m}$.
The first-order condition (at an interior equilibrium) is then $\frac{\dd A_{m}}{\dd q_{m}(n)}=0$, i.e., for all $n$ and $m$,
\begin{equation}
\frac{\partial A_{m}}{\partial
q_{m}(n)}+\sum_{k=1}^N\frac{\partial A_{m}}{\partial p_{k}(m)}\frac{\dd p_{k}%
(m)}{\dd q_{m}(n)}=0,\label{e.FOC_CP_wrt_q}%
\end{equation}%
Recalling that $A_{m}:=\sum_{n}(a-q_{m}(n))t_{n}^{w}e^{-p_n(m)/\theta}$, this gives
\begin{equation}\label{eq:FOC_q}
t_{n}^{w}e^{-p_{n}(m)/\theta}+\sum_{k=1}^N\frac{a-q_m(k)}{\theta}t_k^we^{-p_k(m)/\theta}\frac{\dd p_{k}(m)}{\dd q_{m}(n)}=0.
\end{equation}
Now, let us consider two different ISPs, indexed with $n$ and $n_1\neq n$, and differentiate~\eqref{e.FOC_p_GS} with respect to $q_{m}(n_1)$. 
Using the fact that
$\frac{\dd A_{m}}{\dd q_{m(n_1)}}=0$, we have
\begin{eqnarray*}
\frac{\dd p_n(m)}{\dd q_m(n_1)}&&\hspace{-5ex}\left(1+\underbrace{\frac{1}{\theta}\frac{v}{1-v}\frac{(a-q_{m}(n))t_{n}^{w}e^{-p_{n}(m)/\theta}}{A_{m}}}_{>0}
\right.\times\\&&\times\left.
\underbrace{\left(1-\frac{p_n(m)+q_m(n)}{\theta}\right)}_{>0\text{ from~\eqref{e.FOC_p_GS}}}\right)=0,
\end{eqnarray*}
hence at an equilibrium we have
\begin{equation}
\frac{\dd p_n(m)}{\dd q_m(n_1)}=0\qquad \mbox{ when }n_1\neq n.
\label{e.zero_cross_effects}%
\end{equation}

Now, differentiating~\eqref{e.q+p_nosymmetry} with respect to $q_m(n)$ gives 
\[
\frac{1}{\left(p_n(m)+q_m(n)\right)^2}\left(\frac{\dd p_n(m)}{\dd q_m(n)}+1\right)=0,
\]
that is,\vspace{-3.5ex}
\begin{equation}
\frac{\dd p_n(m)}{\dd q_m(n)}=-1.
\label{e.dp/dq_proven}%
\end{equation}

Plugging~\eqref{e.zero_cross_effects} and~\eqref{e.dp/dq_proven} into~\eqref{eq:FOC_q}, we get
\[
t_{n}^{w}e^{-p_{n}(m)/\theta}\left(1-\frac{a-q_{m}(n)}{\theta}\right)=0,
\]
hence
\begin{equation}\label{e.q_proven}
q_{m}(n)=a-\theta\qquad\mbox{ for all }1\leq m\leq M, 1\leq n\leq N.%
\end{equation}

We remark that those prices are all the same, and are independent of the ISP investments $(t_1,\dots,t_n)$. 

Using the knowledge of all $(q_m(n))$ having the same value $q$, we use~\eqref{e.FOC_p_GS} to re-write $A_{m}$ as
\begin{equation}\label{e.A_new_game_GS}
A_{m}=\theta \frac{v}{1-v} \frac{p_n(m)+q}{\theta-\left(p_n(m)+q\right)} t_{n}^{w}e^{-p_{n}(m)/\theta}\qquad \forall n=1,\dots,N.
\end{equation}

\subsection{ISPs optimal investments $t_{n}$}

Plugging the expression~\eqref{e.A_new_game_GS} for $A_m$ into the expression of ISP investments, we obtain that for any $n,1\leq n\leq N$,
%
\begin{eqnarray*}
\Pi_{T_n}&=&\gamma\left(\theta \frac{v}{1-v}  \frac{v\gamma }{\beta}\right)^{\frac{v}{1-v}}
\times\\&&\hspace{-6ex}\times\!\!
  \sum_{m=1}^{M}\underbrace{t_{n}^{\frac{w}{1-v}}\!\!\left\{\!  (p_n(m)\!+q)^{\frac{1}{1-v}}e^{-\frac{p_{n}(m)}{\theta(1-v)}}\!\!\left(\theta-(p_{n}(m)+q)\right)^{-\frac{v}{1-v}}\!\right\}}_{:= H_m}
\\&&\hspace{-6ex}
-\alpha t_{n}.
\end{eqnarray*}
The first-order condition with respect to investment $t_n$ is
\[
\frac{\partial\Pi_{T_n}}{\partial t_{n}}+\sum\limits_{m}\frac{\partial\Pi_{T_n}}{\partial p_{n}(m)}\frac{\dd p_{n}
(m)}{\dd t_{n}}=0, 
\]
that is, using the notation $K:=\frac{\alpha}{\gamma}\left(\theta \frac{v}{1-v}  \frac{v\gamma }{\beta}\right)^{-\frac{v}{1-v}}$,
\begin{eqnarray}
(1-v)K&\!\!\!=\!\!\!&\sum_{m=1}^M H_m \left(\frac{w}{t_n} + \frac{\dd p_n(m)}{\dd t_n}\left[\frac{1}{p_n(m)+q}-\frac{1}{\theta}
\right.\right.\nonumber\\&&\left.\left.\hspace{16ex}
+\frac{v}{\theta-(p_n(m)+q)}\right]\right)\label{eq:asym_FOC_tn}
\end{eqnarray}


From the symmetry of the model, we will focus on cases when $p_n(m)=p_{n}(m')=p_n$ (i.e., each ISP charges users the same price for all CPs) for any ISP investment profile. Therefore, $\frac{\dd p_n(m)}{\dd t_n}=\frac{\dd p_n}{\dd t_n}$ and all $(H_m)$ have the same value (denoted by $H$), so that~\eqref{eq:asym_FOC_tn} becomes
\begin{equation}\label{eq:semisym_t}
\frac{1-v}{M}\frac{K}{H}=\frac{w}{t_n}+\left(\frac{1}{p_n+q}+\frac{v}{\theta-(p_n+q)}-\frac{1}{\theta}\right)\frac{\dd p_n}{\dd t_n}
\end{equation}

Again from symmetry arguments, we will actually only look here for a {\bf symmetric} equilibrium, where all ISP investments have the same value $\teq$ and all user prices are equal to $p$. To express the first-order conditions, we will look at an infinitesimal deviation in investment from ISP $1$, playing $t_1$ instead of $\teq$, in order to determine the equilibrium common investment value $\teq$.

To use~\eqref{eq:semisym_t}, we therefore need to compute $\frac{\dd p_1}{\dd t_1}$ at our equilibrium symmetric point. This is obtained in two steps:
\begin{itemize}
\item differentiating~\eqref{e.q+p_nosymmetry} with respect to $t_1$ (still at the point ${\bf t}=(\teq,\dots,\teq)$), we obtain (using the fact that $q_m(n)=q$ does not depend on ISP investments) that for all $m$,
\begin{equation}\label{eq:sum_dp_dt1}
\sum_{n=1}^N\frac{\dd p_n(m)}{\dd t_1}=0;
\end{equation}
\item On the other hand,~\eqref{e.A_new_game_GS} and the definition of $A_m$ yield
\[
\sum_{n=1}^Nt_n^we^{-p_n(m)/\theta}=\frac{v}{1-v} \frac{p_{n_1}(m)+q}{\theta-(p_{n_1}(m)+q)} t_{n_1}^{w}e^{-p_{n_1}(m)/\theta}
\]
for any $n_1,1\leq n_1\leq N$. Taking $n_1\neq 1$ and differentiating both sides with respect to $t_1$, we have at our symmetric equilibrium
\begin{eqnarray*}
&&\hspace{-5ex}
\frac{w}{\teq}\teq^we^{-p/\theta}-\frac{1}{\theta}\teq^we^{-p/\theta}\underbrace{\sum_{n=1}^N\frac{\dd p_n(m)}{\dd t_1}}_{=0 \text{ from~\eqref{eq:sum_dp_dt1}}}
\\&&=
N\teq^we^{-p/\theta}\left(\frac{1}{p+q}+\frac{1}{\theta-(p+q)}-\frac{1}{\theta}\right)\frac{\dd p_{n_1}(m)}{\dd t_1},
\end{eqnarray*}
or
$\displaystyle\frac{w}{\teq}=N\left(\frac{1}{p+q}+\frac{1}{\theta-(p+q)}-\frac{1}{\theta}\right)\frac{\dd p_{n_1}(m)}{\dd t_1}.$

Doing the same for all ISPs $n\neq 1$, $\frac{\dd p_{n}(m)}{\dd t_1}=\frac{\dd p_{n_1}(m)}{\dd t_1}$, and therefore from~\eqref{eq:sum_dp_dt1}
\begin{eqnarray}
\frac{\dd p_1(m)}{\dd t_1}&=&-(N-1)\frac{\dd p_{n_1}(m)}{\dd t_1}\nonumber\\
&=&\frac{1-N}{N}\frac{w}{\teq}\frac{1}{\frac{1}{p+q}+\frac{1}{\theta-(p+q)}-\frac{1}{\theta}}\label{eq:sym_diff_dpdt}
\end{eqnarray}
But from~\eqref{e.q+p_nosymmetry}, we obtain 
\begin{equation}\label{eq:p_plus_q}
p+q=\frac{\theta}{1+\frac{v}{N(1-v)}}=\theta\pi,
\end{equation}
with $\displaystyle\pi\define\frac{1}{1+\frac{v}{N(1-v)}},$
\par\noindent which when plugged into~\eqref{eq:sym_diff_dpdt} yields
\begin{equation}
\frac{\dd p_1(m)}{\dd t_1}=-\frac{N-1}{N}\frac{\theta w}{\teq}\frac{1}{\frac{1}{\pi}+\frac{1}{1-\pi}-1}
\end{equation}
\end{itemize}

We can also use~\eqref{eq:p_plus_q} in~\eqref{eq:semisym_t}, and get the equilibrium condition
\begin{eqnarray}
\frac{1-v}{M}\frac{K}{H}&=&\frac{w}{\teq}+\frac{1}{\theta}\left( \frac{1}{\pi}+\frac{v}{1-\pi}-1\right)\frac{\dd p_n}{\dd t_n}\nonumber\\
&=&\frac{w}{\teq}\left(1-R\right)\label{eq:another_condition}
\end{eqnarray}
with $R:=\frac{N-1}{N}\frac{\frac{1}{\pi}+\frac{v}{1-\pi}-1}{\frac{1}{\pi}+\frac{1}{1-\pi}-1}$. Remark that $R\in(0,1)$.

We now need to express the value of $H$, where $p+q=\theta\pi$ from~\eqref{eq:p_plus_q}: 
\begin{eqnarray}
H&=&\teq^{\frac{w}{1-v}}  (\underbrace{p+q}_{\theta\pi})^{\frac{1}{1-v}}e^{-\frac{p}{\theta(1-v)}}(\underbrace{\theta-(p+q)}_{=\theta(1-\pi)})^{-\frac{v}{1-v}}\nonumber\\
&=&\theta\teq^{\frac{w}{1-v}} e^{-\frac{1}{1-v}\left(\pi-q/\theta\right)} \pi^{\frac{1}{1-v}}(1-\pi)^{-\frac{v}{1-v}}\label{eq:express_H}
\end{eqnarray}

Finally, from~\eqref{eq:another_condition} and~\eqref{eq:express_H} we get the expression for the (symmetric) equilibrium ISP investment $\teq$:
\begin{eqnarray}
\teq&=&\left(\left[(1-R)\frac{M}{\tilde K}\frac{w}{1-v} \right]^{1-v}\theta e^{q/\theta-\pi} \pi (1-\pi)^v\right)^{\frac{1}{1-v-w}}\label{eq:t_final}
\end{eqnarray}
\[
\hspace{-9ex}\text{where~~}\left\{
\begin{array}{lcl}
\pi&=&\frac{1}{1+\frac{v}{N(1-v)}}\\
R&=&\frac{N-1}{N}\frac{\frac{1}{\pi}+\frac{v}{1-\pi}-1}{\frac{1}{\pi}+\frac{1}{1-\pi}-1}\\
\tilde K&=&\frac{\alpha}{\gamma}\left(\frac{v}{1-v}  \frac{v\gamma }{\beta}\right)^{-\frac{v}{1-v}}.
\end{array}
\right.
\]
}
\fi
\end{document}

%% file: parameters.tex
\begin{tabular}{ccccccccc}
$a$ & $\theta$ & $w$ & $v$ & $N$ & $\alpha$ & $\beta$ & $c_e$ & $k$\\
\hline
15&10&0.3&0.3&2&1.2&1.2&0.15&0.1\\
\end{tabular}

%% file: M_fixedN_theta.tex
{\footnotesize\begin{tikzpicture}\begin{axis}[legend style={at={(1,1.03)},anchor=south east},width=\figwidth,height=\figheight,cycle list name=\mylist,every axis legend/.append style={nodes={right}},xlabel=$\theta$,ylabel=Number of CPs $M$,legend entries={Non-neutral,Neutral (q=0)},legend columns=2,legend to name=named]
\addplot coordinates{
(6,229)(8.6667,84)(11.333,59)(14,53)(16.667,53)(19.333,57)(22,62)(24.667,69)(27.333,77)(30,86)};
\addplot coordinates{
(6,0)
(8.6667,0)(8.6667,NaN)
(8.6667,32)(11.333,42)(14,50)(16.667,58)(19.333,65)(22,71)(24.667,78)(27.333,84)(30,90)};
\draw (axis cs:10,0) node(aux){};\draw[very thin,opacity=.3] (aux|-current axis.above origin) -- (aux|-current axis.below origin);
\end{axis}\end{tikzpicture}}

%% file: M_fixedN_a.tex
{\footnotesize\begin{tikzpicture}\begin{axis}[legend style={at={(1,1.03)},anchor=south east},width=\figwidth,height=\figheight,cycle list name=\mylist,every axis legend/.append style={nodes={right}},xlabel=$a$,ylabel=Number of CPs $M$,legend entries={Non-neutral,Neutral (q=0)},legend columns=2,legend to name=named]
\addplot coordinates{
(10,0)(10.889,0)
(11.778,0)(11.778,NaN)
(11.778,23)(12.667,34)(13.556,45)(14.444,58)(15.333,73)(16.222,91)(17.111,114)(18,143)};
\addplot coordinates{
(10,0)(10.889,0)(11.778,0)
(12.667,0)(12.667,NaN)
(12.667,22)(13.556,28)(14.444,34)(15.333,39)(16.222,44)(17.111,49)(18,54)};
\draw (axis cs:15,0) node(aux){};\draw[very thin,opacity=.3] (aux|-current axis.above origin) -- (aux|-current axis.below origin);
\end{axis}\end{tikzpicture}}

%% file: t_fixedN_theta.tex
{\footnotesize\begin{tikzpicture}\begin{axis}[legend style={at={(1,1.03)},anchor=south east},width=\figwidth,height=\figheight,cycle list name=\mylist,every axis legend/.append style={nodes={right}},xlabel=$\theta$,ylabel=ISP investments $t$,legend entries={Non-neutral,Neutral (q=0)},legend columns=2,legend to name=named,]
\addplot coordinates{
(6,7.2808)(8.6667,2.6669)(11.333,1.8723)(14,1.6813)(16.667,1.6936)(19.333,1.807)(22,1.9796)(24.667,2.1973)(27.333,2.4521)(30,2.7411)};
\addplot coordinates{
(6,0)
(8.6667,0)(8.6667,NaN)
(8.6667,0.59953)(11.333,1.0244)(14,1.514)(16.667,2.0735)(19.333,2.6987)(22,3.3892)(24.667,4.1448)(27.333,4.9627)(30,5.8423)};
\draw (axis cs:10,0) node(aux){};\draw[very thin,opacity=.3] (aux|-current axis.above origin) -- (aux|-current axis.below origin);
\end{axis}\end{tikzpicture}}

%% file: t_fixedN_a.tex
{\footnotesize\begin{tikzpicture}\begin{axis}[legend style={at={(1,1.03)},anchor=south east},width=\figwidth,height=\figheight,cycle list name=\mylist,every axis legend/.append style={nodes={right}},xlabel=$a$,ylabel=ISP investments $t$,legend entries={Non-neutral,Neutral (q=0)},legend columns=2,legend to name=named,
]
\addplot coordinates{
(10,0)(10.889,0)
(11.778,0)(11.778,NaN)
(11.778,0.73537)(12.667,1.0987)(13.556,1.4525)(14.444,1.8512)(15.333,2.3255)(16.222,2.9083)(17.111,3.633)(18,4.5372)};
\addplot coordinates{
(10,0)(10.889,0)(11.778,0)
(12.667,0)(12.667,NaN)
(12.667,0.56128)(13.556,0.6771)(14.444,0.76314)(15.333,0.82548)(16.222,0.87873)(17.111,0.92582)(18,0.96878)};
\draw (axis cs:15,0) node(aux){};\draw[very thin,opacity=.3] (aux|-current axis.above origin) -- (aux|-current axis.below origin);
\end{axis}\end{tikzpicture}}

%% file: c_fixedN_theta.tex
{\footnotesize\begin{tikzpicture}\begin{axis}[legend style={at={(1,1.03)},anchor=south east},width=\figwidth,height=\figheight,cycle list name=\mylist,every axis legend/.append style={nodes={right}},xlabel=$\theta$,ylabel=CP investments $c$,legend entries={Non-neutral,Neutral (q=0)},legend columns=2,legend to name=named]
\addplot coordinates{
(6,0.064556)(8.6667,0.064464)(11.333,0.064435)(14,0.06441)(16.667,0.064883)(19.333,0.064371)(22,0.064832)(24.667,0.064659)(27.333,0.06466)(30,0.064718)};
\addplot coordinates{
(6,0)
(8.6667,0)(8.6667,NaN)
(8.6667,0.064853)(11.333,0.064564)(14,0.064887)(16.667,0.064349)(19.333,0.064425)(22,0.065093)(24.667,0.064627)(27.333,0.064844)(30,0.064914)};
\draw (axis cs:10,0) node(aux){};\draw[very thin,opacity=.3] (aux|-current axis.above origin) -- (aux|-current axis.below origin);
\end{axis}\end{tikzpicture}}

%% file: c_fixedN_a.tex
{\footnotesize\begin{tikzpicture}\begin{axis}[legend style={at={(1,1.03)},anchor=south east},width=\figwidth,height=\figheight,cycle list name=\mylist,every axis legend/.append style={nodes={right}},xlabel=$a$,ylabel=CP investments $c$,legend entries={Non-neutral,Neutral (q=0)},legend columns=2,legend to name=named]
\addplot coordinates{
(10,0)(10.889,0)
(11.778,0)(11.778,NaN)
(11.778,0.06492)(12.667,0.065613)(13.556,0.065537)(14.444,0.064807)(15.333,0.064683)(16.222,0.064893)(17.111,0.064707)(18,0.064423)};
\addplot coordinates{
(10,0)(10.889,0)(11.778,0)
(12.667,0)(12.667,NaN)
(12.667,0.064632)(13.556,0.065561)(14.444,0.064842)(15.333,0.06491)(16.222,0.064795)(17.111,0.064661)(18,0.064585)};
\draw (axis cs:15,0) node(aux){};\draw[very thin,opacity=.3] (aux|-current axis.above origin) -- (aux|-current axis.below origin);
\end{axis}\end{tikzpicture}}

%% file: UW_fixedN_theta.tex
{\footnotesize\begin{tikzpicture}\begin{axis}[legend style={at={(1,1.03)},anchor=south east},width=\figwidth,height=\figheight,cycle list name=\mylist,every axis legend/.append style={nodes={right}},xlabel=$\theta$,ylabel=User Welfare UW,legend entries={Non-neutral,Neutral (q=0)},legend columns=2,legend to name=named,]
\addplot coordinates{
(6,59.133)(8.6667,21.66)(11.333,15.207)(14,13.655)(16.667,13.755)(19.333,14.677)(22,16.078)(24.667,17.846)(27.333,19.915)(30,22.263)};
\addplot coordinates{
(6,0)(8.6667,4.7962)(11.333,8.1953)(14,12.112)(16.667,16.588)(19.333,21.59)(22,27.114)(24.667,33.158)(27.333,39.702)(30,46.738)};
\draw (axis cs:10,0) node(aux){};\draw[very thin,opacity=.3] (aux|-current axis.above origin) -- (aux|-current axis.below origin);
\end{axis}\end{tikzpicture}}

%% file: UW_fixedN_a.tex
{\footnotesize\begin{tikzpicture}\begin{axis}[legend style={at={(1,1.03)},anchor=south east},width=\figwidth,height=\figheight,cycle list name=\mylist,every axis legend/.append style={nodes={right}},xlabel=$a$,ylabel=User Welfare UW,legend entries={Non-neutral,Neutral (q=0)},legend columns=2,legend to name=named,]
\addplot coordinates{
(10,0)(10.889,0)
(11.778,0)(11.778,NaN)
(11.778,5.9726)(12.667,8.9233)(13.556,11.797)(14.444,15.035)(15.333,18.888)(16.222,23.621)(17.111,29.506)(18,36.85)};
\addplot coordinates{
(10,0)(10.889,0)(11.778,0)
(12.667,0)(12.667,NaN)
(12.667,4.4902)(13.556,5.4168)(14.444,6.1052)(15.333,6.6039)(16.222,7.0298)(17.111,7.4066)(18,7.7502)};
\draw (axis cs:15,0) node(aux){};\draw[very thin,opacity=.3] (aux|-current axis.above origin) -- (aux|-current axis.below origin);
\end{axis}\end{tikzpicture}}

%% file: p_fixedN_theta.tex
{\footnotesize\begin{tikzpicture}\begin{axis}[legend style={at={(1,1.03)},anchor=south east},width=\figwidth,height=\figheight,cycle list name=\mylist,every axis legend/.append style={nodes={right}},xlabel=$\theta$,ylabel=Prices,legend entries={$p$ non-neutral,$p$ neutral,$q$ non-neutral},legend columns=3,legend to name=priceleg,]
\addplot coordinates{
(6,-4.0588)(8.6667,0.80392)(11.333,5.6667)(14,10.529)(16.667,15.392)(19.333,20.255)(22,25.118)(24.667,29.98)(27.333,34.843)(30,39.706)};
\addplot coordinates{
(6,4.9412)(8.6667,7.1373)(11.333,9.3333)(14,11.529)(16.667,13.725)(19.333,15.922)(22,18.118)(24.667,20.314)(27.333,22.51)(30,24.706)};
\addplot coordinates{
(6,9)(8.6667,6.3333)(11.333,3.6667)(14,1)(16.667,-1.6667)(19.333,-4.3333)(22,-7)(24.667,-9.6667)(27.333,-12.333)(30,-15)};
\draw (axis cs:10,0) node(aux){};\draw[very thin,opacity=.3] (aux|-current axis.above origin) -- (aux|-current axis.below origin);
\end{axis}\end{tikzpicture}}

%% file: p_fixedN_a.tex
{\footnotesize\begin{tikzpicture}\begin{axis}[legend style={at={(1,1.03)},anchor=south east},width=\figwidth,height=\figheight,cycle list name=\mylist,every axis legend/.append style={nodes={right}},xlabel=$a$,ylabel=Prices,legend entries={$p$ non-neutral,$p$ neutral,$q$ non-neutral},legend columns=3,legend to name=priceleg,]
\addplot coordinates{
(10,8.2353)(10.889,7.3464)(11.778,6.4575)(12.667,5.5686)(13.556,4.6797)(14.444,3.7908)(15.333,2.902)(16.222,2.0131)(17.111,1.1242)(18,0.23529)};
\addplot coordinates{
(10,8.2353)(10.889,8.2353)(11.778,8.2353)(12.667,8.2353)(13.556,8.2353)(14.444,8.2353)(15.333,8.2353)(16.222,8.2353)(17.111,8.2353)(18,8.2353)};
\addplot coordinates{
(10,0)(10.889,0.88889)(11.778,1.7778)(12.667,2.6667)(13.556,3.5556)(14.444,4.4444)(15.333,5.3333)(16.222,6.2222)(17.111,7.1111)(18,8)};
\draw (axis cs:15,0) node(aux){};\draw[very thin,opacity=.3] (aux|-current axis.above origin) -- (aux|-current axis.below origin);
\end{axis}\end{tikzpicture}}

%% file: Gamma_fixedN.tex
{\footnotesize\begin{tikzpicture}\begin{axis}[legend style={at={(1,1.03)},anchor=south east},width=\figwidth,height=\figheight,cycle list name=\mylist,every axis legend/.append style={nodes={right}},xlabel=$M$,ylabel=$M\gamma$,]
\addplot[thick] coordinates{
(0,0)(3.4359,0.25921)(6.8718,0.54546)(10.308,0.7973)(13.744,1.0093)(17.179,1.1855)(20.615,1.3317)(24.051,1.4539)(27.487,1.557)(30.923,1.6451)(34.359,1.7214)(37.795,1.7885)(41.231,1.8482)(44.667,1.9022)(48.103,1.9515)(51.538,1.9971)(54.974,2.0396)(58.41,2.0795)(61.846,2.1173)(65.282,2.1532)(68.718,2.1875)(72.154,2.2204)(75.59,2.2521)(79.026,2.2827)(82.462,2.3123)(85.897,2.341)(89.333,2.3688)(92.769,2.3958)(96.205,2.4222)(99.641,2.4479)(103.077,2.4729)(106.513,2.4974)(109.949,2.5213)(113.385,2.5447)(116.821,2.5676)(120.256,2.5901)(123.692,2.612)(127.128,2.6336)(130.564,2.6548)(134,2.6755)};
\filldraw[red] (axis cs:67,2.1705) circle (2pt);\draw (axis cs:67,2.1705) node[anchor=south east,red]{non-neutral};\filldraw[blue] (axis cs:37,1.7736) circle (2pt);\draw (axis cs:37,1.7736) node[anchor=north west,blue]{neutral};
\end{axis}\end{tikzpicture}}

%% file: UW_fixedN_v.tex
{\footnotesize\begin{tikzpicture}\begin{axis}[legend style={at={(1,1.03)},anchor=south east},width=\figwidth,height=\figheight,cycle list name=\mylist,every axis legend/.append style={nodes={right}},xlabel=$v$,ylabel=UW,legend entries={Non-neutral,Neutral (q=0)},legend columns=2,legend to name=named,]
\addplot coordinates{
(0.1,6.8721)(0.13333,7.3943)(0.16667,8.239)(0.2,9.4105)(0.23333,11.141)(0.26667,13.627)(0.3,17.353)(0.33333,23.258)(0.36667,33.375)(0.4,52.54)};
\addplot coordinates{
(0.1,2.583)(0.13333,2.6733)(0.16667,2.9159)(0.2,3.3408)(0.23333,3.9887)(0.26667,4.9895)(0.3,6.4219)(0.33333,8.5243)(0.36667,11.814)(0.4,17.536)};
\draw (axis cs:0.3,0) node(aux){};\draw[very thin,opacity=.3] (aux|-current axis.above origin) -- (aux|-current axis.below origin);
\end{axis}\end{tikzpicture}}

%% file: UW_fixedN_c_e.tex
{\footnotesize\begin{tikzpicture}\begin{axis}[legend style={at={(1,1.03)},anchor=south east},width=\figwidth,height=\figheight,cycle list name=\mylist,every axis legend/.append style={nodes={right}},xlabel=CP entry cost $c_e$,ylabel=UW,legend entries={Non-neutral,Neutral (q=0)},legend columns=2,legend to name=named,]
\addplot coordinates{
(0.04,17.407)(0.093333,17.406)(0.14667,17.363)(0.2,17.085)(0.25333,16.351)(0.30667,14.88)(0.36,10.512)(0.41333,0)(0.46667,0)(0.52,0)};
\addplot coordinates{
(0.04,6.8366)(0.093333,6.8053)(0.14667,6.4607)(0.2,5.0973)(0.25333,0)(0.30667,0)(0.36,0)(0.41333,0)(0.46667,0)(0.52,0)};
\draw (axis cs:0.15,0) node(aux){};\draw[very thin,opacity=.3] (aux|-current axis.above origin) -- (aux|-current axis.below origin);
\end{axis}\end{tikzpicture}}